
\magnification=1200
\baselineskip=20pt
\def\lsim{<\kern-2.5ex\lower0.85ex\hbox{$\sim$}\ }
\def\rsim{>\kern-2.5ex\lower0.85ex\hbox{$\sim$}\ }
\overfullrule=0pt
\ \ \
\vskip 3cm
\centerline{\bf COMMENT ON \lq\lq GAUGE--INDEPENDENT ANALYSIS OF}
\centerline{\bf CHERN--SIMONS THEORY WITH MATTER COUPLING"}
\bigskip
\centerline{by}
\bigskip
\centerline{C. R. Hagen}
\centerline{Department of Physics and Astronomy}
\centerline{University of Rochester}
\centerline{Rochester, NY 14606}
\vfil\eject
There have been numerous attempts over the past several years to extend the
anyon concept to the domain of relativistic quantum field theory.  Probably
 the most well known of these is that of Semenoff$^1$ who claimed that upon
 combining a spin zero field operator with an appropriate exponential of
the current operator two desired goals were achieved -- namely, the
statistics of the field became anomalous and the interaction was
 eliminated from the equations of motion.
  Such a result (if correct) would have been
an appropriate basis for claiming a relativistic field theory of anyons.
In fact the necessity of this set of \underbar{two} criteria for an anyon
theory must seem rather evident inasmuch as the anyon view of interacting
nonrelativistic flux tubes succeeded only because it allowed interactions
to be replaced by alterations of the underlying particle statistics.  In
the case considered by Semenoff, however, it was subsequently shown$^{2,3}$
that the claimed elimination of the interaction term required an incorrect
identity, thus leaving unfulfilled the desired goal of an anyon field
 theory.

Recently an approach fairly similar in spirit to that of Semenoff has been
presented$^4$ which claims to circumvent the criticisms of refs. 2 and 3 by
avoiding the use of gauge-fixing.  It is the purpose of this Comment to
point out that (a) there is a serious error in ref. 4 which invalidates one
of its principal claims and (b) independent of that error ref. 4 does not
offer a relativistic field theory of anyons.

The error (a) occurs in the last equation of ref. 4 which essentially
claims that $D_i = \partial_i + i A_i (i = 1,2)$ can be replaced by
$\partial_i$ if one includes a line integral of $A_i$ in the exponential
which defines \lq\lq the gauge-invariant $\hat \phi$".  However, this is
true only if
$$\partial_i \int^x dx \cdot A = A_i (x) \quad ,\eqno(1)$$
an equation which is well known to be generally valid only for the case of
one spatial dimension.  This incorrect result of ref. 4 then leads to the
subsequent claim that \lq\lq explicit dependence on the potential has been
eliminated by the use of careted variables".  This itself is a rather
surprising statement since it has been known ever since this theory was
first formulated$^5$ that it is a \lq\lq photonless gauge theory", i.e., a
model which allows the gauge fields to be written explicitly in terms of
the current operator.  Thus the elimination of explicit reference to the gauge
fields certainly does not require a new (gauge independent or otherwise)
reformulation of the model.

It is worth noting that a definition of the derivative of a path dependent
quantity by means of
$$\partial_i \chi (x,P) = \lim_{dx_i \rightarrow 0} {\chi
(x_i + dx_i, P^\prime) - \chi (x_i,P) \over
dx_i} \eqno(2)$$
with $P^\prime$ an extension of $P$ by $dx_i$ does not remedy the
situation.  This is most easily demonstrated by the counterexample $A = (y,
0)$ which yields for the rhs of (1) the result $x-x_0$ for $i=2$ and the
path choice given by Eq. (10) of ref. 4.  One would obtain the correct
answer for $A_i(x)$ in this case if the $dx_i$ term were absent in the
numerator of (2) so that the entire contribution arose from the difference
between the paths $P^\prime$ and $P$.  However, this further modification
of the derivative would then not reduce to the conventional derivative when
acting on path independent quantities, thereby creating additional
difficulties in the calculations described in ref. 4.

Criticism (b) consists of the observation that even upon ignoring point (a)
ref. 4 is not a relativistic field theory of anyons.  This is actually an
immediate consequence of the first paragraph of this Comment which has
pointed out that it is not enough to define operators which have peculiar
statistics.  That can, of course, always be done in any field theory.  In
fact it is essential that one also \underbar{eliminate} (not merely
rewrite) the effect of the gauge field coupling, leaving only the anomalous
statistics as a residue.  In ref. 4 a rewriting of the equations without
explicit reference to the gauge field
 (actually an invalid claim in view of the criticism (a))
 has been substituted for the
indispensable criterion of the reduction of the scalar field equation to a
Klein-Gordon equation.  Since the latter has not been accomplished in ref.
4 it clearly does not provide a field theory of anyons.

This work was supported by the Department of Energy Grant No.
DE-FG02-91ER40685.

\bigskip
\bigskip

\noindent {\bf References}
\bigskip
\item{1.} G. Semenoff, Phys. Rev. Lett. {\bf 61}, 517 (1988).
\item{2.} C. R. Hagen, Phys. Rev. Lett. {\bf 63}, 1025 (1989).
\item{3.} C. R. Hagen, Phys. Rev. {\bf D44}, 2614 (1991).
\item{4.} R. Banerjee, Phys. Rev. Lett. {\bf 69}, 17 (1992).
\item{5.} C. R. Hagen, Ann. Phys. (N.Y.) {\bf 157}, 342 (1984).
\bye